\documentclass[aps,final,notitlepage,oneside,twocolumn,nobibnotes,nofootinbib,
superscriptaddress,noshowpacs,centertags]{revtex4-1}

\usepackage{graphicx}
\usepackage{latexsym}
\usepackage{amssymb}
\usepackage{amsmath}

\begin{document}

\title{Weighing of the Dark Matter at the Center of the Galaxy}
\author{V.\;I.\;Dokuchaev}
\affiliation{Institute for Nuclear Research of the Russian Academy of Sciences
60th October Anniversary Prospect 7a, 117312 Moscow, Russia}
\affiliation{National Research Nuclear University MEPhI, Kashirskoe sh. 31, Moscow, 115409 Russia}
\author{Yu.\;N.\;Eroshenko}
\affiliation{Institute for Nuclear Research of the Russian Academy of Sciences
60th October Anniversary Prospect 7a, 117312 Moscow, Russia}

\date{\today}

\begin{abstract}
A promising method for measuring the total mass of the dark matter near a supermassive black hole at the center of the Galaxy based on observations of nonrelativistic precession of the orbits of fast S0 stars together with constraints on the annihilation signal from the dark matter particles has been discussed. An analytical expression for the precession angle has been obtained under the assumption of a power-law profile of the dark matter density. In the near future, modern telescopes will be able to measure the precession of the orbits of S0 stars or to obtain a strong bound on it. The mass of the dark matter necessary for the explanation of the observed excess of gamma radiation owing to the annihilation of the dark matter particles has been calculated with allowance for the Sommerfeld effect.
\end{abstract}

\maketitle

\bigskip

Significant advances have been achieved in recent years in the observations of stars gravitationally connected to the supermassive black hole SgrA* at the center of the Galaxy. Several so-called S0 stars, which
move at very high velocities ($>10^3$~km/s) in almost elliptic orbits around a very compact supermassive
object, are observed in the infrared range \cite{Gheetal08,Giletal09-1,Giletal09-2,Meyer12}. Models alternative to the supermassive black hole at the center of the Galaxy, e.g., a cluster of compact stars
such as white dwarfs, neutron stars, or black holes with star masses, are almost excluded \cite{ZeldPodur65,ShaTeu85}. The SgrA* at the center of the Galaxy is most probably a supermassive black hole, although for the final proof it is necessary to confirm the existence of the event horizon of this object.

According to the measured parameters of the Kepler orbits of S0 stars, the mass of supermassive black hole SgrA* is $M_{\rm BH}=(4.1\pm0.4)\times10^6M_\odot$ \cite{Gheetal08,Giletal09-1,Giletal09-2,Meyer12}.
Independent and currently most accurate values of the mass $M_{\rm BH}$ and spin (Kerr parameter) $a$ of the SgrA* black hole are determined from the observations of quasiperiodic oscillations with average periods of 11.5
and 19~min \cite{Aschenbach04,Genzel03}. They are $M_{\rm BH}=(4.2\pm0.2)\times10^6M_\odot$ and $a=0.65\pm0.05$ \cite{dokuch14}.

At the center of the Galaxy, in addition to the supermassive black hole SgrA*, there are additional invisible sources of mass such as compact gas clouds, dim stars and their remnants, and a distributed mass in the form of the dark matter density peak. Constraints on the dark matter density at the center of the Galaxy
based on pulsar effects were discussed in \cite{YuaIok14,BraLin14}. All this additional mass would result in the deviation of the total Newtonian gravitational potential from the potential of a point mass of the black hole $U=-G M_{\rm BH}/r$. As a result, the orbits of S0 stars gravitationally connected to the black hole would be unclosed and precess (see, e.\,g., \cite{LL}). The openness of the orbit of the most studied S0-2 star will be measured in the next one or two years. Thus, the total mass of the dark matter within the orbit of this star with a characteristic radius of $0.005$~pc will be determined. The nonrelativistic precession of orbits of fast S0 stars under consideration, depending on the mass of the dark matter near the center of the Galaxy, can significantly exceed the corresponding relativistic precession (an effect such as the shift of the perihelion of Mercury and frame dragging).

The existence of fast S0 stars provides a unique possibility of reconstructing the gravitational potential
and measuring the mass distribution at the center of the Galaxy by fitting their orbits. The authors of \cite{Gheetal08,Giletal09-1,Giletal09-2} performed a detailed multiparametric fitting of the
orbits of several S0 stars and calculated the additional distributed mass with various exponents of the density
profile. It was shown that the distributed mass within the orbit of the S0-2 star is no more than $3-4$~\% of the mass of the supermassive black hole. It is noteworthy that the expected measurement of the nonrelativistic
precession of the orbit of the S0-2 star will allow either improving the indicated bound on the distributed dark
mass by two or three orders of magnitude or determining this dark mass.  We discuss and develop a method for studying the distribution of the dark matter at the center of the Galaxy by measuring the precession angle of orbits of S0 stars. For a number of particular cases, numerical calculations of the precession angle of orbits of S0 stars because of the extended mass distribution were performed \cite{RubEck01,Mouetal04,Zaketal07,Gualandris10,Zakharov11,Zakharov13}. We obtained general analytical formulas for the precession of orbits of stars with a powerlaw profile of the dark matter; these formulas make it
possible to easily determine the additional distributed mass from the measured precession angle.

\begin{figure}[t]
\begin{center}
\includegraphics[angle=0,width=0.45\textwidth]{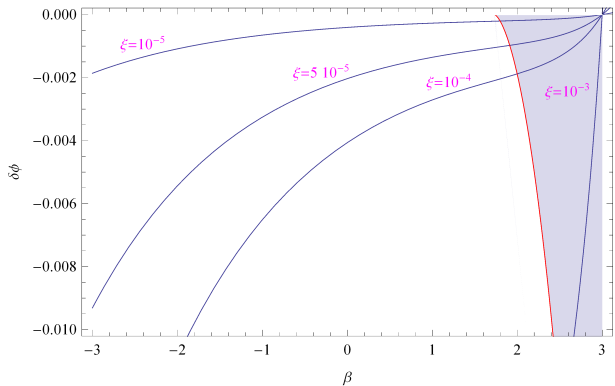}
\end{center}
\caption{Shift angle of the apsis of the orbit of the star in one
turn $\delta\phi$ calculated by Eq.~(\ref{lleq}) versus the exponent of the
power-law spectrum of the dark matter $\beta$ in Eq.~(\ref{power}) for
realistic values of the mass fraction of the dark matter $\xi$
within the orbit of the S0-2 star. The indicated region is
excluded by the constraints caused by the annihilation of
dark matter particles if the dark matter makes the main
contribution to $\xi$. \label{figphi}}
\end{figure}

\begin{figure}[t]
\begin{center}
\includegraphics[angle=0,width=0.45\textwidth]{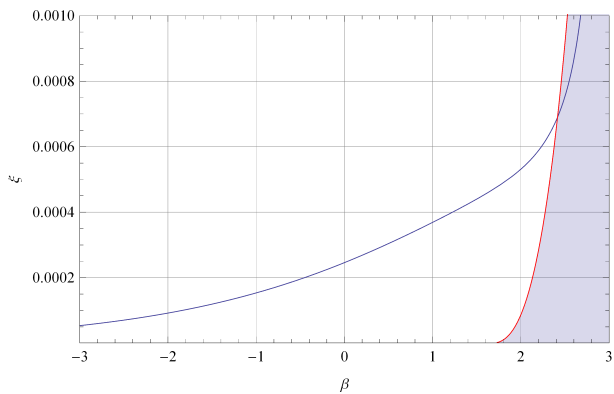}
\end{center}
\caption{Mass fraction of the dark matter $\xi$ versus the exponent $\beta$ in the density profile  given by Eq.~(\ref{power}) at the precession angle $\delta\phi=0.01$. The indicated region is excluded by
the constraints caused by the annihilation of dark matter
particles. \label{figxi}}
\end{figure}

An additional independent method for determining the distribution of the dark matter is the search for
a possible annihilation signal from the center of the Galaxy. The explanation of the excess of a gamma signal with an energy of $\sim1$~TeV from the center of the Galaxy observed by the HESS telescope by gamma
annihilation of dark matter particles with allowance for constraints on the dynamics of stars for the case of
the power-law density profile of the dark matter with a spike and an exponent as a free parameter was analysed
in \cite{HalGon06}. The possibility of constraints on annihilation based on the dynamics of stars or precession was also mentioned in \cite{Zaketal07}. We calculated (see Figs.~\ref{figphi} and \ref{figxi}) the mass of the dark matter necessary for the explanation of the excess of gamma radiation from the center of the Galaxy detected recently by the Fermi-LAT space gamma telescope \cite{Dayetal14,FieShaShe14}. In particular, we determined the dependence of the additional mass both on the profile of the central spike of the dark matter density and on the annihilation cross section of dark matter particles taking into account the Sommerfeld enhancement effect.

In the presence of a small correction $\delta U$ to the Newtonian potential of the black hole, the precession
angle of the orbit of a probe particle (S0-2 star) in one turn is (see \cite{LL}, Sect. 15, Problem 3)
\begin{equation}
\delta\phi=\frac{\partial }{\partial L}\left(\frac{2m}{L}\int\limits_0^\pi \!r^2(\phi)\delta Ud\phi\right)\!.
\label{lleq}
\end{equation}
Here, integration is performed with the trajectory of the particle in the form of an unperturbed elliptic orbit
$r(\phi)=p(1+e\cos\phi)^{-1}$, where $e$ is the eccentricity of the ellipse, $p=L^2/(GM_{\rm BH}m)=a(1-e^2)$ is the parameter of the orbit, a is the major semiaxis, and $L$ is the conserved angular momentum of the star with the mass $m$. The observed parameters of the Kepler orbit of the S0-2 star: the eccentricity $e=0.898\pm0.0034$, the radius of the pericenter $r_p=a(1-e)=0.585$~mpc, and the radius of the apocenter $r_a=a(1+e)=9.42$~mpc. We
note that, in the case of relativistic precession, the orbit would rotate in the direction of the rotation of the star, but Newtonian precession (\ref{lleq}) occurs in the opposite direction, i.e., $\delta\phi<0$.

We consider the power-law density profile of matter responsible for the correction $\delta U$ to the potential of the black hole:
\begin{equation}
\rho(r)=\rho_h\left(\frac{r}{r_h}\right)^{-\beta},
\label{power}
\end{equation}
where $\rho_h$, $r_h$, and $\beta$ are the parameters. The corresponding total mass of the dark matter within the sphere with the radius $r$ is
\begin{equation}
M_{\rm DM}(r)=\frac{4\pi\rho_h r_h^\beta}{3-\beta}\left[r^{3-\beta}-R_{\rm min}^{3-\beta}\right],
\label{mrint}
\end{equation}
where $R_{\rm min}$ is the minimum radius to which the density profile given by Eq.~(\ref{power}) expands. The subsequent calculation of the precession angle of the orbit will be performed under the assumption that $R_{\rm min}<r_p$ and $\beta<3$, i.\,e., that most of the mass of the dark matter within the orbit is located near the apocenter $r=r_a$. We now determine the mass fraction of the dark matter within the orbit of the S0 star $\xi=[M_{\rm DM}(r_a)-M_{\rm DM}(r_p)]/M_{\rm BH}$, which is significant for the subsequent analysis.

The correction to the potential in the case of the power-law profile given by Eq.~(\ref{power}) is
\begin{equation}
\delta U=\left\{
\begin{array}{lll}
\!\!A r^{2-\beta}+\frac{C_1}{r}+C, & \!\!\mbox{åñëè} & \beta\neq2, \label{dpbig}
\\
\!\!4\pi G\rho_h r_h^2m\ln r+\frac{C_2}{r}+C, & \!\!\mbox{åñëè} & \beta=2,
\end{array}
\right.
\end{equation}
where $A=4\pi G\rho_hr_h^\beta m/[(3-\beta)(2-\beta)]$. The constant $C$ does not contribute to the precession angle $\delta\phi$ (because the corresponding contribution to integral (\ref{lleq}) is proportional to $L$) and the term $\propto1/r$ is responsible only for a small addition to the central mass and also does not contribute to the precession angle. The constants $C_1$ and $C_2$ can be represented in the form $C_{1,2}=GmM_{\rm DM}(r_a)$, where $M_{\rm DM}(r_a)$ is the total mass of the dark matter between the event horizon of the black
hole and the radius of the apocenter of the star under consideration.

The calculation of the precession angle of the orbit of the star in the time of one turn around the black hole $\delta\phi$ by Eqs.~(\ref{lleq}) and (\ref{dpbig}) gives an expression with two contiguous hypergeometric functions; with the use of the Gauss relations for contiguous functions, this expression is reduced to the following expression with one hypergeometric function $_2F_1(a,b;c;z)$:
\begin{equation}
\delta\phi=-\frac{4\pi^2\rho_hr_h^\beta p^{3-\beta}}{(1-e)^{4-\beta}M_{\rm BH}}{_2F_1}\left(4-\beta,\frac{3}{2};3;-\frac{2e}{1-e}\right).
\label{itog}
\end{equation}
To test this result, we also calculated the precession angle $\delta\phi$ within standard perturbation theory with the use of the method of osculating elements \cite{Elyasberg}; the resulting expression coincides with Eq.~(\ref{itog}). The precession angle $\delta\phi$ given by Eq.~(\ref{itog}) is negative at all
allowed parameters.

The magnitude of the nonrelativistic precession angle given by Eq.~(\ref{itog}) is in qualitative agreement with
the numerical calculations of precession in  \cite{RubEck01,Mouetal04,Zaketal07,Gualandris10,Zakharov11,Zakharov13}.
Expression (\ref{itog}) for the precession angle at a small eccentricity of the orbit, $e\ll1$, coincides with an accuracy of $e^2$ with the corresponding value calculated analytically by another method in \cite{Iorio}. However, at a large eccentricity, $e\simeq1$, the precession angle calculated in \cite{Iorio} changes sign to positive and diverges in the limit $e\to1$. The formalism used in \cite{Iorio} is possibly
applicable only at $e\ll1$, because the Newtonian precession angle $\delta\phi$ should always be negative.

The function $\delta\phi$ given by Eq.~(\ref{itog}) is continuous at $\beta=2$ (see Fig.~\ref{figphi}). We use Eq.~(\ref{itog}) to perform calculations for various density profiles of the dark matter. We calculate the function $\delta\phi(\beta,\xi)$ in Eq.~(\ref{itog}) and find the level line $\delta\phi(\beta,\xi)= \delta\phi_{\rm obs}$ with the value $\delta\phi_{\rm obs}\sim0.01$ maximum allowable by the observation data (see Fig.~\ref{figxi}). The values $\beta$ and $\xi$ on this line indicate the parameters at which the observation results can be explained.

The excess of gamma radiation from a region with a dimension up to $10^\circ$ from the center of the Galaxy
observed by the Fermi-LAT telescope was fitted (see \cite{Dayetal14,FieShaShe14}) with the generalized Navarro–Frenk–White profile
\begin{equation}
 \rho_{\rm H}(r)=
 \frac{\rho_{0}}{\left(r/d\right)^{\gamma}\left(1+r/d\right)^{3-\gamma}},
 \label{ghalonfw}
\end{equation}
where $d=20$~kpc and $\rho_{\rm H}(8.5\mbox{~kpc})=0.3$~GeV~cm$^{-3}$; the best fit of the Fermi-LAT data is obtained at $\gamma\approx1.26$. Near the center of the Galaxy, at $r\ll d$, this profile is close to the power-law profile
\begin{equation}
 \rho_{\rm H}(r)=\rho_{0} \left(\frac{d}{r}\right)^\gamma.
 \label{ghalonfw-2}
\end{equation}
The central region of the dark matter density peak according to Eq.~(\ref{ghalonfw-2}) is called the cusp. If this profile is directly extrapolated to the center of the Galaxy, the mass of the dark matter within the orbit of the S2 star is
\begin{eqnarray}
M_{\rm DM}&=&M(r_a)-M(R_c)= \nonumber \\
&&\frac{4\pi\rho_0d^\beta(r_a^{3-\gamma}-R_c^{3-\gamma})}{3-\gamma}\simeq2.8M_\odot,
\label{mmmnfw}
\end{eqnarray}
where $R_c={\rm max}\{r_p,r_{\rm ann}\}$, $r_{\rm ann}$ being the possible inner edge of the distribution of the dark matter associated with its annihilation (see below). Quantity (\ref{mmmnfw}) is much smaller than the value accessible for the constraints by the dynamics of stars. However, in the presence of the central black hole, the indicated extrapolation is invalid, because the density profile should be significantly modified by the gravity of the black hole.

The formation of an additional density peak (spike) with the density profile $\propto r^{-\beta}$ around the central black hole was discussed in a number of works \cite{n1,n2}. If the spike was formed adiabatically, i.e., gradually with an increase in the mass of the black hole, the density in the spike could be much higher than the density in the cusp. Fields, Shapiro and Shelton \cite{FieShaShe14} showed that $\beta=2.36$ for the adiabatically formed spike and this spike at $\langle\sigma v\rangle=const$ would be a very bright source at
the center of the Galaxy (see also \cite{BerGurZyb92}). The annihilation signal from such a spike calculated in \cite{FieShaShe14} is a factor of about $\sim35$ stronger than the signal from the extended region with excess of gamma radiation. Since such bright sources at the center of the Galaxy are absent, the existence of the adiabatic spike contradicts observations. It is stated in \cite{FieShaShe14} that the spike could be formed nonadiabatically or be destroyed. In this case, $\beta<2.36$ and the contradiction could be removed.

Following \cite{FieShaShe14}, we write the density of the dark matter in the spike in the form of Eq.~(\ref{power}), where $r_h= GM_{\rm BH}/v_c^2\sim1.7$~pc is the radius of the action of the black hole, $v_c=105\pm20$~km~s$^{-1}$ is the observed standard deviation of velocities at the distance $\sim1$~pc from
the center of the Galaxy, and the density $\rho_{h}$ is determined by matching Eq.~(\ref{power}) with the density given by Eq.~(\ref{ghalonfw}) at the radius $r_h$.

The minimum radius $r_{\rm ann}$ is determined by the annihilation of particles in the time of existence of the
spike (see \cite{FieShaShe14}). This quantity depends on the parameters of particles and the density distribution. According to the calculations in \cite{FieShaShe14}, the best fit of the gamma spectrum is obtained at $m=35$~GeV and $\langle\sigma v\rangle=1.7\times10^{-26}$~cm$^2$s~$^{-1}$; in this case, $r_{\rm ann}\sim {\tilde r}_{\rm ann}\equiv3\times10^{-3}$~pc. It was shown in \cite{FieShaShe14} that contradiction with the bright point source is absent if $\beta=\gamma_s\equiv1.8$. In this case, the mass of the dark matter within the orbit
of the S0-2 star is $M_{\rm DM}\simeq45M_\odot$. Noticeable dynamic effects (precession of the orbit of the star, etc.) should absent at such a small mass.

It was assumed above that $\langle\sigma v\rangle=const$. However, in a number of models of the dark matter $\langle\sigma v\rangle$ can depend on $v$. The velocities $v$ of particles increase when approaching the black hole. The dependence $\langle\sigma v\rangle$ of on $v$ can significantly affect annihilation. Owing to high Kepler velocities near the black hole, the annihilation signal from the center can be reduced, conserving the extended signal from the region $\sim10^\circ$. In particular, $\langle\sigma v\rangle$ depends on $v$ in models involving the Sommerfeld enhancement effect \cite{HisMatNoj04,Pro05,LatSil09}. Sommerfeld enhancement is possible if a dark matter particle is a member of a multiplet of states with close masses, between which coannihilation occurs, e.\,g., in the model of neutralino with the dominance of Higgsino. The gain ${\cal R}$ owing to the Sommerfeld effect is determined from the relation $\langle\sigma v\rangle={\cal R}\langle\sigma v\rangle_0$,
where
\begin{equation}
 \label{SF}
 {\cal R}=\frac{\pi\mu}{b}(1-e^{-\pi\mu/b})^{-1}.
\end{equation}
Here, $\mu=const$ and $b=v/c$. We consider a quite general case where the cross section in the corresponding
region of the parameters can be approximated by the power-law dependence
\begin{equation}
\langle\sigma v\rangle=\langle\sigma v\rangle_0\left(\frac{v_0}{v}\right)^\eta,
\label{powerl}
\end{equation}
where $\langle\sigma v\rangle_0=const$ and $v_0=const$. Power-law dependence (\ref{powerl}) was considered in \cite{BelKirKhl12} in the calculation of the annihilation of the dark matter in self-gravitating bunches. The model with $\langle\sigma v\rangle=const$ and the model with Sommerfeld enhancement at
$\pi\mu/b\ll 1$ correspond to particular cases $\eta=0$ and $\eta=1$, respectively.

The radius $r_{\rm ann}$ at which the maximum density $\rho$ of the dark matter limited by the annihilation effect is reached is determined from the condition
\begin{equation}
n\langle\sigma v\rangle t_g\sim1,
\end{equation}
where $n=\rho/m$ and $t_g\sim10^{10}$~yr is the age of the density peak around the black hole. For cross section (\ref{powerl}), we obtain
\begin{equation}
r_{\rm ann}=r_h\lambda^{\frac{1}{\beta-\eta/2}},
\label{ann1}
\end{equation}
where $\lambda\equiv \rho_h\langle\sigma v\rangle_0 t_g/m$.

\begin{figure}[t]
\begin{center}
\includegraphics[angle=0,width=0.45\textwidth]{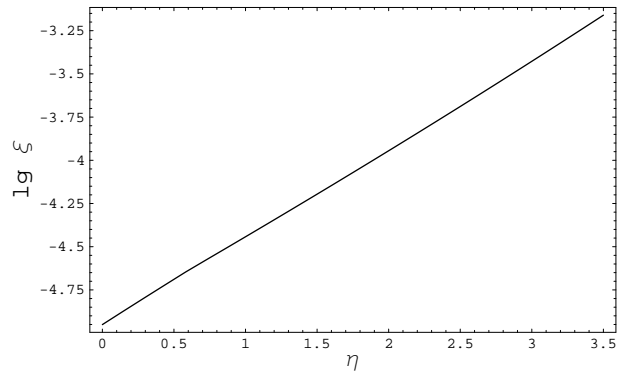}
\end{center}
\caption{Mass fraction of the dark matter $\xi$ required by the
observed excess of gamma radiation in the case of the annihilation of the dark matter particles with the cross section
$\langle\sigma v\rangle\propto v^{-\eta}$ versus $\eta$.). \label{gr3}}
\end{figure}

The velocities of particles near the black hole are $v(r)\approx (GM_{\rm BH}/r)^{1/2}$ at $r<r_h$. Let $3-2\beta+\eta/2<0$. This condition is satisfied for the parameters considered below. The corresponding rate of annihilation of the dark matter in the range of radii from $r_1$ to $r_2$ under the condition $r_1\ll r_2$ can be written in the form
\begin{eqnarray}
\dot N&=&4\pi\int\limits_{r_1}^{r_2}r^2dr\rho^2(r)
m^{-2}\langle\sigma_{\rm ann} v\rangle=
\nonumber
 \\
&=&\frac{4\pi \rho_h^2r_h^{2\beta}\langle\sigma v\rangle_0v_0^\eta r_1^{3-2\beta+\eta/2}}{m^2(GM_{\rm BH})^{\eta/2}(2\beta-3-\eta/2)}.
\label{ann2}
\end{eqnarray}
The authors of \cite{FieShaShe14}, where $\langle\sigma v\rangle=const$ was accepted, found the parameters of the power-law profiles of the cusp and spike that ensure the absence of contradiction with the bright point source at the center. We assume that the integral annihilation signals from the cusp and peak in model (\ref{powerl}) are the same as in \cite{FieShaShe14}. This assumption allows a simple calculation without a detailed fit of the observed excess of gamma radiation. The main signal in the cusp is generated at $r\sim r_h$, where $v\sim v_c$. Hence, fixing the parameter $v_0\equiv v_c$, we obtain $\langle\sigma v\rangle_0 = 1.7\times 10^{-26}$~cm$^2$s~$^{-1}$ as in \cite{FieShaShe14}. The density profile in the peak in our case differs from that used in \cite{FieShaShe14} because of the dependence $v(r)$. We determine the density profile in the peak at $\langle\sigma v\rangle\propto v^{-\eta}$ taking into account the above assumptions. Equating the annihilation rate given by Eq.~(\ref{ann2}) in the spike at $\beta=\gamma_s=1.8$ and $\langle\sigma v\rangle=\langle\sigma v\rangle_0$ to corresponding rate (\ref{ann2}) at arbitrary values $\beta$ and $\langle\sigma v\rangle=\langle\sigma v\rangle_0v_0^\eta/v^{\eta}$, we obtain the nonlinear equation
\begin{equation}
\frac{x}{x+3-\beta}\ln\lambda+\ln(\varkappa x)=0,
\label{nlineq}
\end{equation}
where
\begin{equation}
\beta=\frac{3}{2}+\frac{\eta}{4}+\frac{x}{2}, \quad
\varkappa=\frac{1}{2\gamma_s-3}\left(\frac{r_h}{\tilde r_{\rm ann}}\right)^{2\beta_s-3},
\label{varkappa}
\end{equation}
and $\lambda$ is determined from Eq.~(\ref{ann1}). We solve numerically Eq.~(\ref{nlineq}) with respect to $x$ and, then, calculate the mass of the dark matter by Eq.~(\ref{mmmnfw}). The results of the calculations are shown in Fig.~\ref{gr3}. The equality $r_{\rm ann}=r_p$ is reached at $\eta=0.6$ and the inequality $r_{\rm ann}<r_a$ is always valid under the accepted conditions. Thus, in Eq.~(\ref{mmmnfw}), $R_c=r_p$ at $\eta<0.6$ and $R_c=r_{\rm ann}$ at $\eta>0.6$. The adiabatic density profile $\beta=2.36$ is reached at $\eta=3.13$.

In particular, the mass of the dark matter $M_{\rm DM}$ within the orbit of the S0-2 star in particular cases $\eta=0$, $1$, $3.13$, and $3.5$ is $45M_\odot$, $144M_\odot$, $1.8\times10^3M_\odot$, and $2.8\times 10^3M_\odot$, respectively. These values correspond to the values $\xi=1.2\times10^{-5}$, $3.6\times10^{-5}$, $4.4\times10^{-4}$, and $6.9\times10^{-4}$ and $\beta=1.8$, $1.9$, $2.36$, and $2.4$, respectively.
These values are upper bounds on the possible $\xi$ and $\beta$ values. In the first two cases, the distributed mass of the dark matter is still too small to affect dynamic effects (see Fig.~\ref{gr3}). At the same time, the real prospect of the measurement of the additional mass of the dark matter from the precession of S0 stars appears already at $\eta>3$.

The currently existing observation accuracy is still insufficient for the measurement of the precession
angle of fast S0 stars and the distributed invisible mass. However, there is a high probability of reaching in the near future the accuracy required either for the measurement of the precession angle or for the determination of a strong bound, which in turn will make it possible to impose stringent dynamic constraints on the additional dark mass. If the invisible mass is attributed to annihilating particles, the observation of the annihilation signal from the center of the Galaxy provides additional possibilities for the calculation of the distributed mass or for the determination of bounds on it. According to Fig.~\ref{gr3}, at $\langle\sigma v\rangle=const$ and even with Sommerfeld enhancement $\langle\sigma v\rangle\propto1/v$, the dynamics of stars still cannot give constraints on annihilation, because the mass of the dark matter within the orbit of the S0-2 star in these cases is very small. At the existing accuracy, the dynamics of stars and annihilation are independent. However, if the annihilation cross section depends on the velocity with a large exponent $\eta>3$ in Eq.~(\ref{powerl}), the mass of the dark matter can be significant. In this case, joint constraints could be obtained in the near future from the dynamics of stars and from the data on gamma radiation from the center of the Galaxy.

This work was supported by the Division of Physical Sciences, Russian Academy of Sciences (program no. OFN-17); by the Russian Foundation for Basic Research (project no. 13-02-00257); and by the Leading Scientific
Schools grant (project no. NSh-3110.2014.2).



\begin{thebibliography}{9}

\bibitem{Gheetal08}
A.\,M.\,Ghez, S.\,Salim, N\,N.\,Weinberg, et al., Astrophys. J. \textbf{689}, 1044 (2008).

\bibitem{Giletal09-1}
S. Gillessen, F. Eisenhauer, S. Trippe, et al., Astrophys. J. \textbf{692}, 1075 (2009).

\bibitem{Giletal09-2}
S. Gillessen, F. Eisenhauer, T. K. Fritz, et al., Astrophys. J. \textbf{707}, L114 (2009).

\bibitem{Meyer12}
L. Meyer, A.\,M., Ghez, R. Sch\"odel, et al. Science, \textbf{338}, 84 (2012).

\bibitem{ZeldPodur65} Ya.\,B.\,Zel'dovich, M.\,A.\,Podurets,
Astron. Zhurnal, \textbf{42}, 963 (1965); Soviet Astron. \textbf{9}, 742 (1966).

\bibitem{ShaTeu85} S.\,L.\,Shapiro, S.\,A.\,Teukolsky, Astrophys. J. Lett \textbf{292}, L41 (1985).

\bibitem{Aschenbach04}
B. Aschenbach, N. Grosso, N. Porquet, P. Predehl, Astron. Astrophys. \textbf{417}, 71 (2004).

\bibitem{Genzel03}
R. Genzel, R. Sch\"odel, T. Ott, et al. Nature, \textbf{425}, 934 (2003).

\bibitem{dokuch14} V.\,I. Dokuchaev, Gen. Relativ. Gravit. \textbf{46}, 1832 (2014).

\bibitem{YuaIok14}
Q. Yuan, K. Ioka, arXiv:1411.4363 [astro-ph.HE].

\bibitem{BraLin14}
J. Bramante, T. Linden, Phys. Rev. Lett. \textbf{113}, 191301 (2014).

\bibitem{LL} L.\,D.\,Landau and E.\,M.\,Lifshitz, Course of Theoretical Physics, Vol. 1: Mechanics (Fizmatlit, Moscow, 2004; Pergamon Press, New York, 1988), ch. 3.

\bibitem{Elyasberg} P.\,E.\,El'yasberg,  Introduction to the Theory of Flight of Artificial Earth Satellites. Israeli Program for the Translation of Scientific Publications, Jerusalem. 1967.

\bibitem{RubEck01}
G.\,F. Rubilar and A. Eckart, Astron. Astrophys. \textbf{374}, 95 (2001).

\bibitem{Mouetal04}
N.\,Mouawad, A.\,Eckart, S.\,Pfalzner, et al., Astron. Nachr. \textbf{326}, 83 (2005).

\bibitem{Zaketal07}
A.\,F.\,Zakharov, A.\,A.\,Nucita, F.\,De\,Paolis et al., Phys. Rev. \textbf{D 76}, 062001 (2007).

\bibitem{Gualandris10}
A. Gualandris, S. Gillessen, D. Merritt, Mon. Not. Roy. Astron. Soc. \textbf{409}, 1146 (2010).

\bibitem{Zakharov11}
F. de Paolis, G. Ingrosso, A.\,A. Nucita, et al., Gen. Rel. Gravit. \textbf{43}, 977 (2011).

\bibitem{Zakharov13}
D. Borka, P. Jovanovic, V. Borka Jovanovi\'c, A.F. Zakharov, JCAP \textbf{11}, 050, (2013).

\bibitem{Iorio}
L. Iorio, Galaxies, \textbf{1}, 6, (2013).

\bibitem{HalGon06}
J. Hall, P. Gondolo, Phys. Rev. \textbf{D 74}, 063511 (2006).

\bibitem{Dayetal14}
T. Daylan, D. P. Finkbeiner, D. Hooper, et al., arXiv:1402.6703 [astro-ph.HE].

\bibitem{FieShaShe14}
B. D. Fields, S. L. Shapiro, J. Shelton, Phys. Rev. Lett. \textbf{113}, 151302 (2014).

\bibitem{BerGurZyb92}
V.\,S.\,Berezinsky, A.\,V.\,Gurevich, K.\,P.\,Zybin, Phys. Lett. \textbf{B 294}, 221 (1992).

\bibitem{HisMatNoj04}
J. Hisano, S. Matsumoto,  M.\,M.\,Nojiri, Phys. Rev. Lett. \textbf{92}, 031303 (2004)

\bibitem{Pro05}
S. Profumo, Phys. Rev. \textbf{D 72}, 103521 (2005).

\bibitem{LatSil09}
M. Lattanzi, J. Silk, Phys. Rev. \textbf{D 79}, 083523 (2009).

\bibitem{n1} P. Gondolo, J. Silk, Phys. Rev. Lett. 83 1719 (1999).

\bibitem{n2} O.Y. Gnedin, J.P. Primack, Phys. Rev. Lett. 93 061302 (2004).

\bibitem{BelKirKhl12}
K.\,M.\,Belotsky, A.\,A.\,Kirillov, M.\,Yu.\,Khlopov, Gravit. Cosmol. \textbf{20}, 47 (2014).

\end{thebibliography}
\end{document}